# Temperature limits in laser cooling of free atoms with three-level cascade transitions

Flavio C. Cruz*, Michael L. Sundheimer, and Wictor C. Magno

*Departamento de Fisica, Universidade Federal Rural de Pernambuco, Recife, PE, 52171-900, Brazil*
*\*Instituto de Fisica Gleb Wataghin, Universidade Estadual de Campinas, Campinas, SP, 13083-859, Brazil*

We employ semiclassical theoretical analysis to study laser cooling of free atoms using three-level cascade transitions, where the upper transition is much weaker than the lower one. This represents an alternate cooling scheme, particularly useful for group II atoms. We find that temperatures below the Doppler limits associated with each of these transitions are expected. The lowest temperatures arise from a remarkable increase in damping and reduced diffusion compared to two-level cooling. They are reached at the two-photon resonance, where there is a crossing between the narrow and the partially-dark dressed states, and can be estimated simply by the usual Doppler limit considering the decay rate of the optical coherence between these states.

## I. INTRODUCTION

Laser cooling is an extremely successful technique, which has caused a profound impact in several fields of physics [1, 2]. Providing a high degree of control of atomic quantum systems, its applications range from increased precision in atomic spectroscopy and the development of atomic clocks [3], the formation and studies of quantum degenerate gases [4], and ultracold molecules [5], studies of solid-state physics with atoms trapped in optical lattices [6], to cooling of micromechanical resonators [7]. The basic Doppler cooling mechanism of momentum transfer from a light beam in near resonance with a two-level, freely moving atom, leads to a well-known temperature limit (typically several hundred μKelvin for metal-alkaline atoms [2]) that is proportional to the scattering rate (linewidth) of the atomic transition and given by the balance between friction (cooling) and diffusion (heating) in the atom-laser interaction [8], [9]. Lower temperatures are achieved using techniques which exploit the multilevel structure of atoms combined with optical pumping mechanisms, such as in polarization gradient cooling [2], velocity-selective coherent population trapping (VSCPT [10]), Raman cooling [11], or sideband cooling in the case of confined atoms or ions [12].

Such sub-Doppler or sub-recoil techniques in particular cannot be applied to the most abundant spinless isotopes of two-electron atoms, such as group IIa, IIb and also Yb, whose ground states lack hyperfine or Zeeman structure. Minimum temperatures were therefore limited by the linewidth of their strong resonant $^1S_0$-$^1P_1$ cooling transition. A successful solution was to employ a subsequent stage of Doppler cooling using the narrower intercombination transition, a scheme that was originally adopted for trapped ions [13]. Although this solution worked well for heavier atoms such as Sr [14] or Yb [15], it is difficult to be applied for lighter elements, such as Ca or Mg, for which the intercombination transition is a "clock" one, and has too small of a scattering rate. This required quenching the transition by coupling it to an upper level with faster decay rate [16], [17]. A last scheme involved the opposite situation, in which the lower transition is broad, or strongly allowed (such as cooling transitions in alkaline atoms), and the upper is narrower or weaker [18], [19], [20]. Cooling is obtained when they are simultaneously excited by two lasers in an EIT (electromagnetically induced transparency) fashion, e.g., by a weak ("probe") and a strong ("dressing") laser, which respectively drive the strong and the weak transitions. This scheme was experimentally demonstrated for magnesium [21], [22] and cesium [23], and showed cooling below the Doppler limit of the strong transition. An extension of the technique was also proposed for cooling of anti-hydrogen [24]. Despite experimental and theoretical work [18-23], the question of what is the Doppler limit for laser cooling with three-level cascade transitions still needs to be clarified. Previous analysis [18], [20] stated that minimum temperatures would approach the limit given by the narrower of the two transitions. This intuitive result is expected for large single-photon detunings, since in this case the intermediate state could be adiabatically eliminated, leading to an effective two-level system composed of the lower and the upper states.

In this paper we reexamine this basic cooling scheme, extending the analysis of two-level cooling, showing that temperatures below the Doppler limits given by any of the individual transitions are expected. These "sub-Doppler" limits are reached at the two-photon resonance, including the Stark shift. In the dressed state picture, this point corresponds to a crossing of the narrow and (partially) dark states. We verified that the Doppler limit can be simply calculated from the decay rates of these two dressed states. We use standard laser cooling analysis and the dressed state picture to identify and discuss different cooling regions, such as on the blue side of the two-photon resonance [23], and to set optimum parameters for experiments. Although motivated by cooling of alkaline-Earth atoms, our results can also be applied and tested in metal-alkaline atoms, or extended to the theory of cooling with femtosecond lasers [25]. We note that cooling based on quantum interference and dark resonances has been previously used in VSCPT [10] and EIT cooling [26]. VSCPT was originally implemented with Λ-transitions on free-atoms [10], using two beams of equal intensities and frequencies. Atoms are optically pumped into a stationary dark state at near zero velocity, leading to sub-recoil temperatures. EIT cooling has been demonstrated using Λ-transitions in trapped ions [27], excited by a strong "dressing" and a weak "probe" laser, leading to cooling to the ground state of the trap. By using both lasers blue-detuned from their respective resonances, and adjusting the Rabi frequency of the dressing laser, the spectrum can be tailored such that the red-sideband transition is enhanced while the carrier and blue-sideband transitions are largely suppressed.

Cooling based on 3-level transitions has also been studied in [28], [29], again considering Λ-transitions in trapped ions. Ref. [29] has focused on the temperature limit, finding that it should depend on the linewidth of the two-photon transition (which can be very small for a Λ-transition with long-lived lower states) or the ratio of Rabi frequencies. The scheme analyzed here is similar to EIT cooling, but we focus on free atoms having cascade transitions with dissimilar wavelengths and widths, particularly in which the lower transition is broad and the upper is narrow. This is the opposite situation of quench cooling demonstrated with the intercombination transition of alkaline-Earth atoms [16, 17], in which the lower transition is much narrower than the upper one.

## II. EXPERIMENTAL AND THEORETICAL DEFINITIONS

The experimental configuration involves one-dimensional optical molasses, with counterpropagating pairs of laser beams. We neglect interference effects for the light, collective effects for the atoms, and assume a homogeneously broadened transition, as for two-electron atoms already pre-cooled with the resonant transition to a few mKelvin. Laser polarizations can be adjusted in order to induce 2-photon absorptions from co- or counterprogating beams (for example, by using equal or opposite circular polarizations for copropagating beams). This would also imply in different recoil limits, for transitions with different wavelengths. We have considered realistic experimental situations using atomic parameters for group IIa, IIb, and rare-Earth (Yb) atoms. An example of a suitable cascade cooling transition is the $^1S_0$-$^1P_1$-$^1D_2$, where the lower $^1S_0$-$^1P_1$ transition is the usual cooling one. Figure 1 shows the level diagram, with the nomenclature that we use in the calculations considered in this paper.

Throughout this paper we will use, as an example and without loss of generality, the numerical values for this transition in magnesium ($\lambda_1 = 2\pi c/\omega_1 = 285$ nm, $\lambda_2 = 2\pi c/\omega_2 = 881$ nm, $\gamma_1 = 2\pi(78.8$ MHz$)$, $\gamma_2 = 2\pi(2.2$ MHz$)$, since this element is one that seems to benefit most from this cooling scheme, and for which experiments have been done. We considered excitation with laser intensities corresponding to Rabi frequencies $\Omega_1 = 2\pi(10$ MHz$) \sim 0.12\gamma_1$ and $\Omega_2 = 2\pi(80$ MHz$) \sim 36.36\gamma_2$, or saturation parameters $S_1$ and $S_2$ [$S_i = 2(\Omega_i/\gamma_i)^2$] equal to 0.034 and 2645, respectively. Our analysis is based on well-known semiclassical treatments of laser cooling [30], [31].

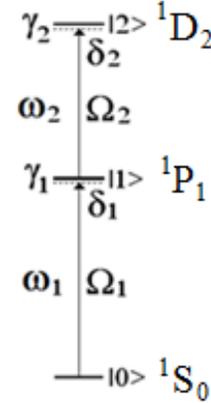

Figure 1. Ladder three-level system considered in this paper, with terms indicating suitable cooling transition in alkaline-Earth atoms. Lasers 1 and 2 have Rabi frequencies $\Omega_1$ and $\Omega_2$, and detunings $\delta_1$ and $\delta_2$. Levels 2 and 1 decay by spontaneous emission with rates $\gamma_2$ and $\gamma_1$.

The time evolution of the density matrix operator $\hat{\rho}$ is described by the Liouville equation, including the relaxation terms:

$$\frac{d\hat{\rho}}{dt} = \frac{1}{i\hbar}\left[\hat{H},\hat{\rho}\right] + \Gamma\hat{\rho} \quad . \quad (1)$$

The Hamiltonian operator $\hat{H}$ in the bare states basis and using the rotating wave approximation is given by [19]:

$$\hat{H} = \hat{H}_0 + \hat{V}_{AL} = \frac{\hbar}{2}\cdot\begin{pmatrix} 0 & \Omega_1 & 0 \\ \Omega_1 & -2\delta_1 & \Omega_2 \\ 0 & \Omega_2 & -2\delta_1 - 2\delta_2 \end{pmatrix}, \quad (2)$$

where $\hat{H}_0$ represents the unperturbed Hamiltonian and $\hat{V}_{AL}$ is the atom-laser interaction potential.

We solve the Liouville equation in steady-state regime using the rotating-wave approximation. The atomic velocity is introduced via the detunings $\delta_{1,2} - \mathbf{k}_{1,2}\cdot\mathbf{v}$ (k is the wavenumber). The force due to the interaction of the atoms with the light fields can be calculated by:

$$\hat{F} = -\nabla\hat{V}_{AL}(\vec{R}) \quad , \quad (3)$$

$$F = \langle\hat{F}\rangle = Tr[\hat{F},\hat{\rho}] \quad , \quad (4)$$

where $F$ represents the expectation value of the force operator described in the center-of-mass motion of the atom at position $\vec{R}$. Considering only the radiation pressure force corresponding to the two counter-propagating incident light beams, we then have:

$$F = \frac{-i\hbar}{2}\{k_1\Omega_1[\rho_{01}(\delta_1+k_1\cdot\mathbf{v},\delta_2+k_2\cdot\mathbf{v})-\rho_{10}(\delta_1+k_1\cdot\mathbf{v},\delta_2+k_2\cdot\mathbf{v})]$$
$$+ k_2\Omega_2[\rho_{12}(\delta_1+k_1\cdot\mathbf{v},\delta_2+k_2\cdot\mathbf{v})-\rho_{21}(\delta_1+k_1\cdot\mathbf{v},\delta_2+k_2\cdot\mathbf{v})]$$
$$- k_1\Omega_1[\rho_{01}(\delta_1-k_1\cdot\mathbf{v},\delta_2-k_2\cdot\mathbf{v})-\rho_{10}(\delta_1-k_1\cdot\mathbf{v},\delta_2-k_2\cdot\mathbf{v})]$$
$$- k_2\Omega_2[\rho_{12}(\delta_1-k_1\cdot\mathbf{v},\delta_2-k_2\cdot\mathbf{v})-\rho_{21}(\delta_1-k_1\cdot\mathbf{v},\delta_2-k_2\cdot\mathbf{v})]\}$$
. (5)

This semiclassical analysis allows the interpretation in terms of the damping ($\alpha$) and diffusion coefficients (D), whose balance sets the effective temperature of the system $T = D/\alpha k_B$. The damping coefficient is the derivative of the radiation pressure force with respect to velocity [Eq.(5)]. Since the force is proportional to the transition scattering rate and atomic velocity is proportional to detuning, the damping coefficient is related to the derivative of the atomic lineshape. The diffusion coefficient, which sets the minimum temperature, reflects the momentum diffusion due to light scattering and is also calculated from the force [30]. We verified that the correlation term in the diffusion coefficient is negligible compared to the spontaneous one [30, 31]. Our curves for temperature, damping or diffusion coefficients have been normalized by the Doppler limit for cooling only on the lower transition ($\delta_1=-\gamma_1/2$, and $S_1=0.001$, leading to the Doppler limit $T_1 = \hbar\gamma_1/2k_B = 1.9$ mK for this Mg transition). We have first verified our calculations considering cooling with the lower two-level transition only. We also verified that the estimated temperatures are well above the recoil limit of the individual transitions in the cascade system, in order to avoid a limit in which semiclassical theories should not be valid. We note that the semiclassical analysis is not meaningful near the two-photon resonance for the bare states, where the damping coefficient goes to zero.

## III. EFFECTIVE TEMPERATURE AND DRESSED STATES DECAY RATES

In Figure 2 we show a contour plot of effective temperature as function of single-photon detunings $\delta_1$ and $\delta_2$, for the Mg transition discussed in this paper. Regions of significant cooling are seen for negative and positive single-photon detunings, and also on the blue side of the two-photon resonance for the bare states (above the diagonal given by $\delta_1+\delta_2=0$). Figure 3 shows a plot of temperature as function of $\delta_2$, with $\delta_1=-1.96\gamma_1$. A minimum temperature of $0.02T_1=38\mu K$ is obtained for $\delta_1=-1.96\gamma_1$ and $\delta_2=1.82\gamma_1$, which corresponds to the two-photon resonance considering the Stark shift, $\delta_2=-(\delta_1+\delta_{Stark})$ with $\delta_{Stark} = \frac{1}{2}\left(\sqrt{\delta_2^2+\Omega_2^2}-\delta_2\right)$. This limit is lower than the Doppler limit that would be associated with an effective two-level transition between states $|0\rangle$ and $|2\rangle$, $T_{2D} = \hbar\gamma_2/2k_B = 53\mu K$. A less pronounced cooling on the blue side of the two-photon resonance is also predicted, and has been observed in previous experiments in Mg [21], [22] and Cs [23].

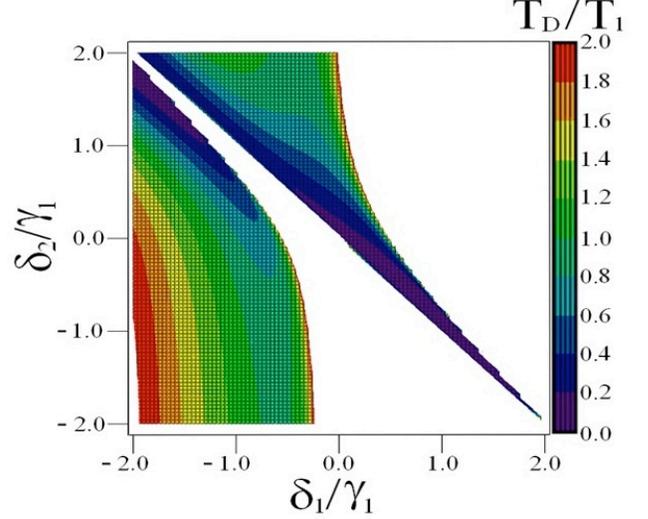

**Figure 2.** (Color online) Contour plot of temperature (normalized to the Doppler limit $T_1 = \hbar\gamma_1/2k_B$), as function of single-photon detunings $\delta_1$ and $\delta_2$. $^1S_0$-$^1P_1$-$^1D_2$ three-level cascade transition in Mg, Rabi frequencies are $\Omega_1 = 2\pi(10$ MHz$)$ and $\Omega_2 = 2\pi(80$ MHz$)$.

An useful insight into the cooling mechanism can be gained by analyzing the curves for the damping and diffusion coefficients, $\alpha$ and D [2], plotted in Figure 4. The curve for D has exactly the same shape as the curve for the absorption coefficient or the population of the intermediate state. The absorption coefficient reflects the resonances of the dressed atom, showing a broad resonance, a narrow resonance and a minimum of absorption close to the two-photon resonance, corresponding to a (partially) dark resonance. In the dressed state picture, the atom can be in three eigenstates, linear combinations of the bare states.

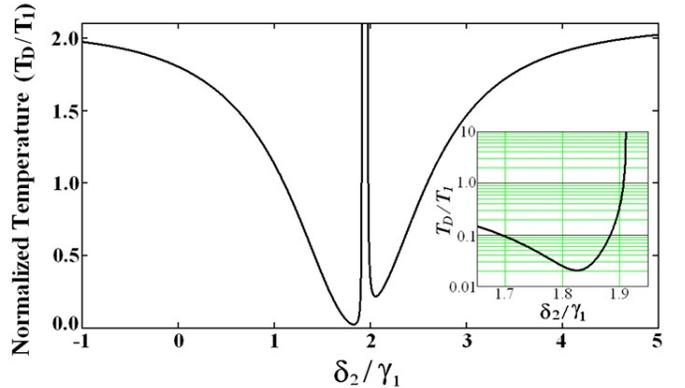

**Figure 3.** Calculated temperature for cooling with the $^1S_0$-$^1P_1$-$^1D_2$ three-level cascade transition in Mg, normalized to the Doppler limit of the lower transition, $T_1 = \hbar\gamma_1/2k_B$ (obtained at $\delta_1 = -\gamma_1/2$ and for $\Omega_1 = 2\pi(10$ MHz$)$). Parameters are $\Omega_1 = 2\pi(10$ MHz$)$, $\Omega_2= 2\pi(80$ MHz$)$ and $\delta_1 = -1.96 \gamma_1$.

The dressed states are eigenstates of the Hamiltonian and their energies and linewidths are calculated as the real and imaginary parts of the eigenvalues λ of Eq. (2) [32], [33]. At the two-photon resonance, they are given by [34]:

$$|\psi_+\rangle = \cos\theta|1\rangle + \sin\theta|\psi_c\rangle \qquad , \qquad (6)$$

$$|\psi_-\rangle = \sin\theta|1\rangle - \cos\theta|\psi_c\rangle \qquad , \qquad (7)$$

$$|\psi_D\rangle = \frac{1}{\Omega}(\Omega_2|0\rangle - \Omega_1|2\rangle) \qquad , \qquad (8)$$

where $|\psi_c\rangle = \frac{1}{\Omega}(\Omega_1|0\rangle + \Omega_2|2\rangle)$, $tg\theta = \frac{1}{\Omega}\left(\sqrt{\delta_1^2 + \Omega^2} - \delta_1\right)$, and $\Omega = \left(\sqrt{\Omega_1^2 + \Omega_2^2}\right)$.

Since for the conditions of Fig.4, with $\Omega_2 \gg \Omega_1$, and $\delta_2 = 1.82\gamma_1$, the "broad" ($|\psi_+\rangle$) and "narrow" ($|\psi_-\rangle$) states are dominated by states $|1\rangle$ and $|2\rangle$, with small contributions from the others, we can think of those resonances as approximate transitions to these bare states ($|0\rangle \to |1\rangle$ and $|0\rangle \to |2\rangle$). Thus the broad resonance can be associated with cooling mostly on the lower transition, while the narrow resonance is associated with cooling on the two-photon transition. In Fig. 4 a remarkable increase of 30 times is seen in the damping coefficient associated with the narrow transition, in comparison with the broad transition (two-level cooling). This can be interpreted considering the higher slope of this resonance, which leads to a higher radiation pressure force. The cooling and damping times, inversely proportional to α [9], should be reduced by the same amount. The diffusion coefficient D has a simple analytical expression, $D = 2\hbar^2(k_1^2\gamma_1\rho_{11} + k_2^2\gamma_2\rho_{22})$ in which the second term is negligible compared to the first, since $\gamma_2 \ll \gamma_1$ and $k_2 < k_1$. We can see explicitly that the curve for D reflects the population of the intermediate state ($\rho_{11}$). The diffusion coefficient is also reduced (by 60%, Fig. 4) with respect to the broad resonance, as a result of the much smaller scattering in the lower transition.

These results indicate that the minimum temperatures occur at the two-photon resonance (including the Stark shift, $\delta_1 = -(\delta_2 + \delta_{Stark})$) where the damping coefficient is maximum, differently of two-level cooling [9]. This is not necessarily the point of maximum slope in the curve of the absorption coefficient [19]. By looking indirectly at this curve in Fig.4 (e.g., through the curve for D), we can identify the different cooling regions in Fig.2 as usual Doppler cooling on the red side of the broad and narrow resonances, with the corresponding limits given approximately by $\gamma_1$ and $\gamma_2$. It is not surprising that the plot in Fig.2 resembles the dressed state diagram for the three-level system (see for example Fig.9 in [34]). In particular, cooling on the red side of the narrow dressed resonance corresponds to cooling on the blue side of the bare two-photon resonance, as experimentally observed [21], [22], [23].

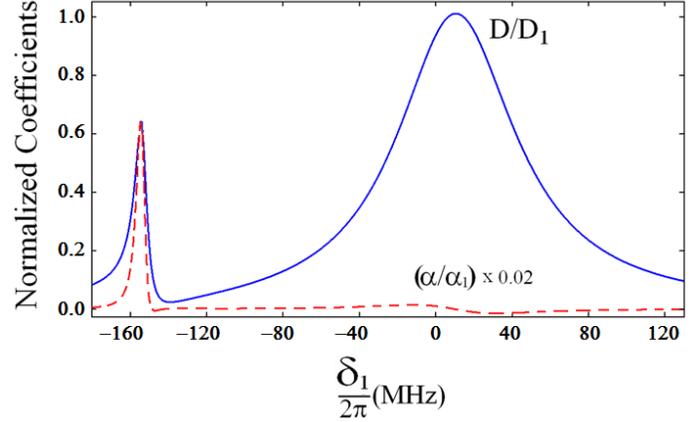

**Figure 4.** (Color online) Damping (dashed curve) and diffusion (solid curve) coefficients for the $^1S_0$-$^1P_1$-$^1D_2$ three-level cascade transition in Mg, normalized to the respective values, $\alpha_1$ and $D_1$, for optimum cooling on the lower $^1S_0$-$^1P_1$ transition (at $\delta_1 = -\gamma_1/2$). As in Fig.3, $\Omega_1 = 2\pi(10\ MHz)$, $\Omega_2 = 2\pi(80\ MHz)$ and $\delta_2 = 1.82\ \gamma_1$.

Figure 5 plots the energies and decay rates (normalized by $\hbar\gamma_1$) of the dressed states for the same parameters of Fig. 3. In Fig. 5a, it is seen that $|\psi_D\rangle$ crosses $|\psi_-\rangle$ at the two-photon resonance, where there is a small change in their decay rates (Fig. 5b). At this point, the population in $|\psi_D\rangle$ is maximized and in $|\psi_-\rangle$ is minimized (the dressed state populations are $\rho_{--} = 11.3\%$, $\rho_{++} = 0.32\%$ and $\rho_D = 88.3\%$, and the bare state populations are $\rho_{00} = 90.1\%$, $\rho_{11} = 0.94\%$ and $\rho_{22} = 8.92\%$). The state $|\psi_D\rangle$ is not completely dark, due to a small contribution from the upper state $|2\rangle$, and has a decay rate quite smaller than $\gamma_2$. If the probe and dressing cooling lasers are phase-locked, at the crossing the atom can be in a superposition of $|\psi_D\rangle$ and $|\psi_-\rangle$. We verified that the temperature at this point can be obtained by the usual Doppler cooling limit [9], considering the effective linewidths of these two dressed states, $\gamma_{eff} = (\gamma_D + \gamma_-)/2$, or in other words, the decay rate of the optical coherence between them. The limit $T_D = \hbar\gamma_{eff}/2k_B$ agrees exactly with the limit obtained in Fig.3. This represents a significant simplification in the estimation of the temperature limit, since the linewidths are easily calculated from the Hamiltonian of Eq. (2), without the need to solve the optical Bloch equations. We note that this result resembles the one obtained in [29], in the analysis for cooling with a Λ-transition, where the lowest temperature was proportional to the decay rate of the two-photon optical coherence.

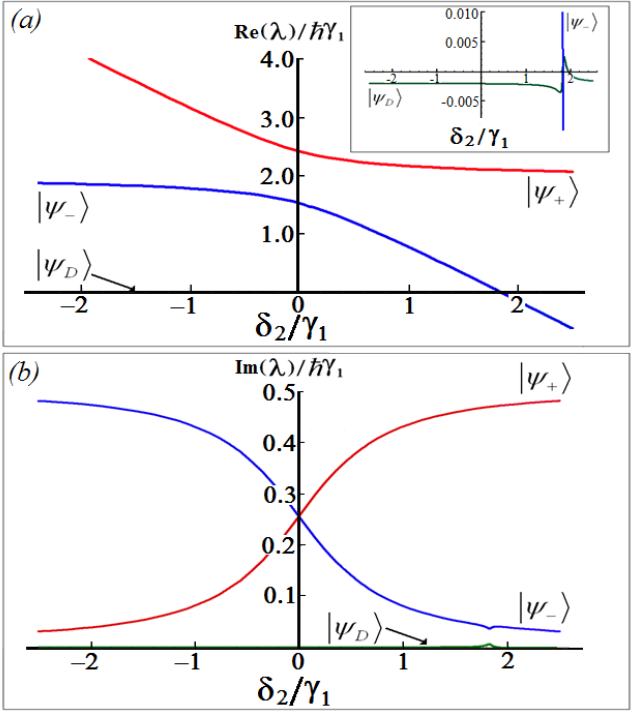

**Figure 5.** (Color online) (a) Energy level diagram for the dressed states [$\delta_1 = -1.96\,\gamma_1$, $\Omega_1 = 2\pi(10\text{ MHz})$, $\Omega_2 = 2\pi(80\text{ MHz})$]. Inset: crossing of the dark and narrow states at the two-photon resonance. (b) Corresponding decay rates (linewidths) of the dressed states. Note the change at the crossing, where the Doppler limit can be estimated by the average of both widths.

## IV.  DISCUSSION

From the above discussion, one can summarize a guide for optimum cooling with three-level cascade transitions, where the lower transition is broad and the upper is narrow. The detuning of the probe laser should be set to $\delta_1 > \gamma_1$, such that the population in $|1\rangle$ is small. Too much detuning $\delta_1$, on the other hand, will require increasing intensities for the dressing laser.

The intensities of both lasers should be adjusted such that $S_1 \ll 1$ and $S_2 \gg 1$ ($\Omega_2 \gg \Omega_1$). Also $\delta_2$ should be adjusted to the two-photon resonance, taking into account the Stark shift, $\delta_2 = -(\delta_1 + \delta_{Stark})$. The minimum temperature will be reached at this condition, given by the Doppler limit $T_D = \hbar(\gamma_D + \gamma_-)/4k_B$.

A few comments can be extended to other three-level systems, composed of transitions with dissimilar linewidths. Similar results should be expected for Λ-type systems that include the ground state ($|0\rangle$), as long as the other lower state ($|2\rangle$), connected to the upper one ($|1\rangle$) by the dressing laser, is not very long lived, such as a hyperfine level of the ground state, such as in typical EIT experiments. This would produce a dark state with negligible decay rate, but also a narrow dressed state with too small of a decay rate (if the rate $\gamma_{12}$ is also small), making three-level cooling very inefficient. This seems to be the situation discussed in [29], in which the two-photon coherence (between $|0\rangle$ and $|2\rangle$) was explicitly assumed to have a finite lifetime. For V-type systems, the "dark" state will be a linear combination of the two upper bare states, with a larger contribution from the broader one, and therefore with no advantages for additional cooling by coupling to a narrower transition.

## V.  CONCLUSION

We have analyzed laser cooling of free atoms using three-level cascade transitions, where the upper transition has a much smaller linewidth than the lower one. We have used standard semiclassical analysis of laser cooling and the dressed state picture. Cooling is expected at the red side of the broad and narrow resonances of the dressed states, with minimum temperatures approaching the Doppler limits given by the linewidths of the bare states. We found however that the minimum temperature can be lower than these, and is achieved at the two-photon resonance, where there is a crossing of the partially dark and narrow dressed states. The minimum temperature is then given by the decay rate of the optical coherence between these dressed states. The familiar expression for two-level cooling can be used to estimate the Doppler limit. Our analysis did not include effects such as interference between the laser beams, or polarization gradients. In practice, heating due to laser intensity fluctuations is also known to cause a temperature limit higher than the Doppler limit [35]. Due to the role of coherences involving three atomic states and high laser intensities, it might be necessary to phase-lock both cooling lasers in order to properly implement the technique experimentally, which has not been done so far. Further experimental tests can be implemented more easily in metal-alkaline atoms.

This work has been supported by FAPESP, FACEPE, CNPq, and CEPOF.
* Corresponding author email: flavio@ifi.unicamp.br